\documentstyle[12pt,aasms4]{article}

\Large
\def\phx{Phoenix}
\def\etal{et al.}

\def\teff{{\rm T}_{\rm eff}}
\def\te{{\rm T}_{\rm eff}}

\def\msol{M_\odot}
\def\mso{M_\odot}

\slugcomment{Refereed}

\begin{document}
\begin{sf}
\bibliographystyle{apj}

\title{\sf Synthetic Spectra and Mass Determination of the Brown Dwarf Gl229B}

\author{\Large\sf France Allard}
\affil{Dept.\ of Physics, Wichita State University,\\ 
Wichita, KS 67260-0032\\
E-Mail: \tt allard@eureka.physics.twsu.edu}

\author{\Large \sf Peter H. Hauschildt}
\affil{Dept.\ of Physics and Astronomy, Arizona State University, Box 871504,\\
Tempe, AZ 85287-1504\\
E-Mail: yeti@sara.la.asu.edu}

\author{\Large \sf Isabelle Baraffe and Gilles Chabrier}
\affil{C.R.A.L. (UMR 142 CNRS), Ecole Normale Sup\'erieure,\\
69364 Lyon Cedex07, France\\
E-Mail: ibaraffe, chabrier @cral.ens-lyon.fr}

\section{ \sf ABSTRACT}
\bigskip

We present preliminary non-grey model atmospheres and interiors
for cool brown dwarfs.  The resulting synthetic spectra are compared 
to available spectroscopic and photometric observations of the coolest 
brown dwarf yet discovered, Gl229B (Nakajima \etal\ \markcite{nak95} 1995).  
Despite the grainless nature of the present models, we find the resulting 
synthetic spectra to provide an excellent fit to most of the spectral 
features of the brown dwarf.  We confirm the presence of methane absorption 
and the substellar nature of Gl229B. These preliminary models set an upper 
limit for the effective temperature of 1000~K.  We also compute the evolution 
of brown dwarfs with solar composition and masses from 0.02 to 0.065 
$\msol$.  While uncertainties in the age of the system yield some 
undetermination for the mass of Gl229B, the most likely solution is 
$m\approx 0.04-0.055 \msol$. In any case, we can set an upper limit 
$m= 0.065\msol$ for a very unlikely age $t=10$ Gyr.


\section{ \sf About Gl229B}
\bigskip

A systematic survey of the nearby stars led by the Caltech-JHU 
collaboration (Nakajima \etal\ \markcite{nak95} 1995) has recently 
achieved a breakthrough in the search for brown dwarfs.  With the 
Johns Hopkins University Adaptive Optics Coronograph, D.A. Golimowski
and B.R. Oppenheimer uncovered the coolest brown dwarf ever found,  
Gl229B, that is 10 times fainter than previously known brown dwarf 
candidates, and is orbiting an M1 type dwarf only 5.7~pc from the 
Earth.  Spectra of the object obtained on the Hale 200-inch 
telescope by Oppenheimer \etal\ \markcite{opp95} (1995) and by a 
number of other investigators reveal unique spectral signatures 
reminiscent of those seen in Jupiter.  If confirmed, the detection 
of methane bands in Gl229B could establish beyond doubt the substellar 
nature of the object. 

Recent work by Tsuji \etal\ \markcite{tsuji96a}\markcite{tsuji96b} 
(1996a, b) has revealed the possible importance of grain formation 
upon the spectroscopic properties of very low mass stars and brown 
dwarfs. In this paper, we show how the various spectroscopic features 
of Gl229B can only be reproduced by {\bf grainless} homogeneous models, 
suggesting that grains must have either sunk below the photosphere 
over the brown dwarf's lifetime, or must be clumped in clouds as 
suggested by Tsuji \etal\ \markcite{tsuji96b} (1996b).  Despite the 
proximity of the system, the large orbital separation will not permit 
a dynamical measure of the mass for at least a decade to come.  We 
therefore use our grainless models to set a meaningful upper limit 
on the effective temperature and mass for the brown dwarf.

\section{\sf Brown Dwarf Model Atmospheres}
\bigskip

We have constructed a grid of non-grey model atmospheres in radiative, 
convective and local thermodynamic equilibrium using the stellar 
atmosphere code \phx. Molecular and atomic opacities are treated 
line-by-line with Van der Waals pressure broadening. Convection is 
included through the mixing length approximation. The models are 
characterized by the following parameters: (i) surface gravity, 
sampling $\log \,g =$ 4.0 to 5.8, (ii) effective temperature, $\teff 
\unskip$, here taken from 1600 to 900$\,$K in steps of 100~K, (iii) 
the ratio $l$ of mixing length to scale height, here taken to be unity, 
(iv) micro-turbulent velocity $\xi$, here set to $\xi=2\,{\rm kms}^{-1}$, 
and (v) the elemental abundances, which we set to solar values (Grevesse 
\& Noels \markcite{abund} 1993).  Inhomogeneities such as the formation 
of clouds and effects of condensation and grain opacities are not included
in the models at this point.  

This grid provides an extension of the red dwarf work by Allard \etal\ 
\markcite{h2olet} (1994) and Allard \& Hauschildt \markcite{MDpap} (1995, 
hereafter AH95) to the brown dwarfs realm.  Pertinent additions to \phx\   
since AH95 include: (1) an upgrade of the equation of state (EOS) from 98 
to 206 molecules using the polynomial partition functions by Irwin 
\markcite{irwin88} (1988); (2) the addition of H$_3^+$ and H$_2^+$ from 
Neale \& Tennyson \markcite{h3p} (1995) and Neale \etal\ \markcite{h3pl} 
(1996) to both the EOS and the opacities; (3) the use of the TiO line list 
from J{\o}rgensen \markcite{TiOJorg} (1994) and (4) the H$_2$O line list 
from Miller \etal\ \markcite{h2olist} (1994); the inclusion of (5) IR 
rotational collision induced absorption of H$_2$ (CIA) by Zheng \& Borysow 
\markcite{ciarot} (1994) and (6) the IR roto-vibrational system of SiO by 
Langhoff \& Bauschlicher \markcite{sio93} (1993); and finally, (7) the 
incorporation of the Hitran and Geisa molecular transitions data banks for 
31 molecules including CH$_4$ and NH$_3$ from Rothman \etal\ 
\markcite{hitran92} (1992) and Husson \etal\ \markcite{geisa} (1992) 
respectively.  

The limitations of the Hitran and Geisa data banks with respect to methane 
have been discussed in detail by Strong \etal\ \markcite{strong} (1993). 
These data bases were originally intended for Earth atmosphere applications 
and contain only the strongest (about 47,000) of the millions of transitions 
expected for this molecule under stellar gas conditions which would provide 
a pseudo-'continuum' for the bands.  We nevertheless include these in the 
present preliminary models as a diagnostic of the presence of methane bands 
in the atmospheres.  More accurate models will require a calculation of the 
CH$_4$ molecule from first principles.  To date, only one such calculation 
exists, which lists CH$_4$ lines only redward of about 3~$\mu$m (Tyuterev 
\etal\ \markcite{CH4Hil} 1994).

\section{\sf Spectral Analysis}
\bigskip

\subsection{\sf Temperature and surface gravity of Gl229B}

The most reliable way of determining the effective temperature of a cool
dwarf is by fitting its overall spectral distribution.  In Fig. 1, we 
compare the latest infrared broadband fluxes by T. Nakajima (IAU Circular 
6280) to our models at four temperatures between 1600 and 900~K (i.e.
surface temperatures of 700 to 400~K) at a constant gravity of log~g= 5.0. 
Water vapor, which only condenses from the gas phase at about 300-373~K, 
dominates the entire infrared spectral distribution in Fig.~1.  Note, however,
that methane absorption breaks through the water continuum in the H and K 
bandpass, and near 3.4~$\mu$m at effective temperatures below 1600~K (the 
upper range of our models).  Since a star at the hydrogen burning limit 
has an effective temperature of about 2000~K (Baraffe \etal\ 
\markcite{baraffe95} 1995), methane which is particularly obvious in the L' 
band spectral region therefore clearly reveals the substellar nature of Gl229B.
We draw particular attention to the spectral range from about 3.5 to 5~$\mu$m, 
where four bandpasses sample a region over which our models are especially 
temperature sensitive.  The 4.3 to 5.2~$\mu$m CO bands dominate this spectral 
region in cool M~dwarfs and ``hot'' brown dwarfs, disappearing at lower 
temperatures as CH$_4$ and CO$_2$ gradually form at the expense of CO.  
Therefore, cooler brown dwarf models ($\teff < 1000$~K) are left with a much 
broader peak around 4.5~$\mu$m.  The observed L' and N band fluxes indicate 
that no CO aborption is seen in Gl229B, and 1000~K is the hottest model which 
lies within the N-band error bar, regardless of gravity.  Furthermore, the 
fact that the M and N fluxes are still above our coolest model suggests that 
Gl229B is even cooler than 900~K and more flux redistribution is needed to 
match these bandpasses.  However, it is possible that these fluxes are 
still affected by uncertainties in the calibration.  

In Fig.~2, we compare our 1000~K and 900~K models to the Hale spectrum of 
Gl229B, recently recalibrated by Matthews \etal\ \markcite{gl229b} (1996).  
The locations of all the major absorption features of Gl229B are relatively 
well reproduced by the models.  The agreement is better in spectral ranges 
dominated by water i.e. in the J bandpass, and in the blue wing of the K 
bandpass.  For the temperatures occuring in the atmospheres of brown dwarfs 
($400$~K~$\le$~T~$\le\ 2000$~K in this model grid), the H$_2$O opacities 
adopted here (as prescribed by Schryber \etal\ \markcite{schryb94} 1995)
appear adequate to describe the data.  The intrinsic temperature sensitivity 
of the water opacity profile in this spectral range results in a strong 
response of the K bandpass flux to changes in the atmospheric structure,
causing the models to be quite sensitive to gravity in the K bandpass flux.  
We find that the K/J flux ratio can equally be satisfied by temperature and 
gravity pairings of $\teff=$1000~K;$log\,g=$5.3 (upper panel of Fig.~2) or 
900~K;$log\,g=$4.8 (lower panel of Fig.~2). Yet the width of the J bandpass 
flux, which is shaped by H$_2$O opacities, is best reproduced by the former
solution.  Missing methane transitions and flux calibration uncertainties in 
the H and K bandpass can easily account for the remaining discrepancies and 
amount to a 0.2 dex error in each gravity estimate. 

\subsection{\sf The role of grains and condensation in GL229B}

Our solar mix models predict strong absorption bands of VO and FeH bluewards 
of 1.1~$\mu$m which are not observed in the spectrum of Gl229B (see Fig.~2).  
This could be an indication of photospheric heating by grains (Tsuji \etal\ 
\markcite{tsuji96a} 1996a) which tend to bring the spectral distribution 
closer to a simple thermal radiation, i.e. a black body.  In such case, 
Gl229B would have to be much cooler than derived with our grainless models 
in order to exhibit the observed water and methane features. This would imply 
an age for the system in conflict with the present estimate (see \S 5 below).
Effectively, a parallel investigation by Tsuji \markcite{1996b} (1996b) failed 
to reproduce the spectral distribution of Gl229B using homogeneous dusty 
models, leading the authors to suggest that rather than dust being distributed 
homogeneously with the gas in the brown dwarf regime it gathers in dust clouds.
Another possibility is that grains could sink below the photosphere while the 
shrinking convection zone is no longer able to return this material to the 
surface.  

In either case, this leads to reduced photospheric abundances as observed
in Gl229B {\bf without} the expected greenhouse effect of grains upon the 
photosphere.  To gauge the effects of condensation on our models and the 
atmospheric parameters we infer from them, we have recomputed the 
($\teff=1000$~K;log~g=5.3) model in the limiting case of Jovian abundances 
(C/O = 1.9 and nearly zero metallicity).  While we found that decreased 
metallicity did improve the fit to the H bandpass and the $\lambda < 1.1~\mu$m
region of the Gl229B spectrum, it also slightly lowers the K band flux 
relative to the J bandpass.  Thus, a bluer J-K color and slightly higher 
gravity model are required to fit the spectrum, an effect which is however 
negligible compared to present uncertainties in the gravity estimates.  

To be conservative, we set upper limits to the effective temperature and 
surface gravity of Gl229B at $\teff = 1000$~K (based on the L' to N fluxes)
and log~g$ = 5.3 \pm 0.2$ (based on the relative K and J fluxes).

\section{\sf Evolution and Mass Determination}
\bigskip

As brown dwarfs of increasing mass are more compact, hotter and older 
at a given luminosity, the effective temperature and gravity derived in 
the previous section can help limit the allowed ranges of mass and age 
for Gl229B.  We have computed the time evolution of brown dwarfs from 
0.02 $\mso$ to 0.065 $\mso$ with solar metallicity.  Details of the 
input physics for the internal structure can be found in Baraffe \etal\ 
\markcite{baraffe95} (1995) and Chabrier \etal\ \markcite{CBP96} (1996). 
The atmospheric structure is defined for temperatures down to $\teff = 
900$~K by the models already described in \S 3. 

We determined a maximum temperature for Gl229B of $\teff \sim 1000$~K.  
The latest estimate of its bolometric luminosity, corrected for the 
scattered light of the primary, using broadband photometry out to 
about 10~$\mu$m, and using a blackbody tail extrapolation to longer 
wavelengths is $L \sim 6.4 \times 10^{-6}$~L$_{\odot}$ (Matthews \etal\ 
\markcite{gl229b} 1996).  Also, the properties of the primary Gl229A can 
be used to constrain the age of the binary (cf. Nakajima \etal\ 
\markcite{nak95} 1996).  Its kinematics suggest that the dM1 star belongs 
to the young disk population, even though the absence of H$_\alpha$ emission 
excludes a very young age.  However, the detection of H$_\alpha$ absorption, 
which we were able to confirm in the Keck spectra of Gl229A by G. Marcy and 
G. Basri (private communication), may also rule out a very old object (see 
e.g. Reid, Hawley and Mateo \markcite{reid95} 1995).  This is supported by 
the metal-rich properties deduced from photometry (cf. Leggett 
\markcite{legg92} 1992).  Independently, the photometric properties of the 
primary Gl229A (M$_V$ = 9.33 and V-I = 2.0, I-K = 1.96, cf. Leggett 
\markcite{legg92} 1992) are accurately reproduced by a $\sim$ 0.6 $\mso$ 
star as modelled by Chabrier \etal\ \markcite{CBP96} (1996).  Between 0.5 
- 5 Gyrs, the main sequence location of a 0.57 $\mso$ star with solar 
metallicity corresponds to M$_V \sim 9.30$ and V-I = 1.90, I-K = 1.94.  We 
thus restrict our study to a time interval of $0.5 - 5$ Gyrs.

Fig.~3 shows the evolution of the effective temperature for different 
masses as a function of age.  In the age range specified above, the only 
masses compatible with the temperature limit set by the synthetic spectral 
analysis lie between 0.02 and 0.055 $\mso$. The high sensitivity of the 
gravity to the mass at a given temperature can be used to further narrow 
the mass range.  In Fig.~4 we plot the variation of the surface gravity 
$g=GM/R^2$ as a function of $\teff$ for the same masses as in Fig.~3.  The 
best fits from the spectral analysis are also shown in this figure with an
uncertainty on the gravity of $\pm 0.2$ (c.f. \S 4.2).  For the temperature 
limit, $\teff$ = 1000~K and $log \, g = 5.3 \pm 0.2$, our favored choice 
(see \S 4.1), the mass range is now reduced to 0.04 - 0.055 $\mso$. We also 
show our results for the pairing $\teff$ = 900K; $log \, g = 4.8 \pm 0.2$ 
which yields 0.02 - 0.035 $\mso$.  An examination of Fig.~3 and 4 reveals 
that a gravity of $log \, g < 4.5$ with $\te\, = 900$ K would yield too 
young ages ($\tau <5\times 10^8$ yrs) for the system.  For completeness, 
the characteristics of our model brown dwarfs at each mass in the entire 
grid with $\te$ = 1000 and 900~K are given in Table~1.  Note that the
observed important blueshift, due to CH$_4$ opacities in the K bandpass,
characteristic of the transition ``from the stellar to the sub-stellar'' 
domain is qualitatively reproduced by the models.

In order to estimate the effect of grain formation on the brown dwarf 
cooling, we have recomputed the evolutionary models based on a grey 
atmosphere using the Alexander and Ferguson \markcite{alex94} (1994) 
opacities with and without grains (Alexander, private communication).  
As shown by Burrows \etal\ \markcite{burws93} (1993) and Chabrier \etal\ 
\markcite{CBP96} (1996), the luminosity drops substantially when the 
location of grain formation reaches the bottom of the photosphere.  
Therefore, the grainless model will evolve at higher L and $\te$.  
We find that at a given age, the differences between models with and 
without grains are $\sim$ 100~K in $\te$ and $\sim$ 25\% in L in the 
temperature range of interest.  This is illustrated in Fig.~3 for the 
0.045~$\mso$ model.  Note that, for the afore-determined temperature 
range, grainless non-grey models predict a similar evolution as grey 
models including grain formation, and thus yield similar mass 
determination (see e.g. Burrows et al. 1993, Figure 4). Though the effect 
of grain formation on evolution might be more important when using non-grey 
model atmospheres, the present analysis based on grey atmosphere models 
changes the results by only $\sim 0.01\,\msol$.

\section{\sf Discussion and Concluding Remarks}
\bigskip

We have used preliminary grainless model atmospheres to model the 
spectral distribution of cool brown dwarfs and found that these 
reproduce well the observed spectroscopic and photometric 
properties of Gl229B.   We confirm the presence of methane 
absorption bands and therefore the substellar nature of Gl229B.
The absence of CO absorption in the 4-5~$\mu$m range sets a secure 
upper limit of 1000~K on the effective temperature.  Gravity can be 
constrained at fixed temperature using the relative K to J band fluxes 
and this leads to values ranging from log~g=5.3 at 1000~K to log~g= 4.8 
at $\teff\ = 900~K$.  A minimum temperature and gravity can be set by 
the presence of water vapor bands in Gl229B's spectrum to about 700-800~K 
where photospheric temperatures are expected to drop below the condensation 
temperature of H$_2$O. 

While uncertainties in the age of the system and in the temperature 
of Gl229B yield some indetermination for the mass of Gl229B, we can 
derive upper limits from the effective temperature limit, namely 
$M\le 0.065 \mso$ for an age $t\le$10 Gyr, and $M\le 0.055 \mso$ for 
a more likely age $t\le$5 Gyr.  The most likely solution $M\approx 
0.04-0.055$ $\msol$ is supported by the quality of the most reliable 
fit to the water bands for these parameters, and would be consistent 
with an age similar to that of our solar system.  Our analysis cannot 
however exclude a mass as low as 0.02 $\mso$ if $\te = 900$ K and 
$t=0.5$ Gyr.

One of the main sources of uncertainty in the present models is the 
absence of grain formation in the computed atmospheres. Recent work 
by Tsuji \etal\ \markcite{tsuji96a}\markcite{tsuji1996b} (1996a,b) shows 
that grain formation may lead to substantial heating of the photospheres 
of cool M dwarfs and brown dwarfs, possibly resulting in a much lower 
effective temperature for Gl229B than derived with the present grainless 
models.  The absence of predicted VO and FeH features in the red spectrum 
(Z band) of Gl229B indicates reduced abundances of grain forming elements,
and therefore the presence of condensation and perhaps grain heating in 
this cool brown dwarf.  However homogeneous dusty models by Tsuji \etal\ 
\markcite{tsuji96b} (1996b) failed to reproduce the spectral distribution
of Gl229B while the present models reproduce the observed strengths of 
molecular features such as water vapor which would otherwise be substantially 
weakened by the greenhouse effect of grains.  This suggests that Gl229B's 
atmosphere is very reminiscent of that of Jupiter, in which condensed 
material and cloudtop layers were not found in the expected amounts by the 
Galileo atmospheric probe (D. Isbell and D. Morse, NASA press release 
96-10).  Perhaps in both Jupiter and Gl229B, these condensates either 
formed in clouds or sank deeper into the atmosphere.

We thank T. Nakajima and colleagues for providing updated version of the 
Gl229B spectrum in electronic form as well as the referee for his valuable
comments.  We are also indepted to J. M. Matthews and D. R. Alexander for 
proofreading the text.  This research has been partially supported by a NASA 
LTSA and ATP grants to ASU and an NSF grant AST-9217946 to WSU.  The 
atmospheric calculations have been performed on the Cray C90 of the San 
Diego Supercomputer Center and on the IBM SP2 of the Cornell Theory Center, 
supported by the NSF.

\clearpage

\message{\the\textheight}

\begin{table}[htbpe]
\caption{Characteristic of brown dwarfs for [M/H]=0}
\begin{tabular}{lllllll}\hline
$M/\msol$  & $\te$ &  age (Gyrs) & L/L$_\odot$
&  log g  & M$_K$ & {J-K\tablenotemark{*}} \\ \hline
  0.02   &1000 & 0.37 & -4.97 & 4.66 & 14.51 & 0.75 \\
  - & 900 & 0.522 & -5.17& 4.67 & 15.36 & 0.44\\
0.025&1000 & 0.61 & -5.01 & 4.80 & 14.69 & 0.64 \\
  - & 900 & 0.88 & -5.22 & 4.82 & 15.58& 0.29\\
0.035 & 1000& 1.43 & -5.10 &  5.02 & 15.07 &  0.42 \\
 -  & 900 & 1.99 & -5.27 &  5.04 & 15.87 &  0.09 \\
0.04 & 1000 & 2.03 & -5.13 &  5.11 & 15.22 &  0.33 \\
  -  &  900 & 2.81 & -5.31 &  5.13 & 16.02 & -0.01 \\
 0.045&1000 & 2.74 & -5.16 & 5.19 & 15.33 & 0.25\\
 - & 900 & 3.87 & -5.35 & 5.21 & 16.21  & -0.12  \\
 0.055&1000 & 4.88 & -5.21 & 5.34 & 15.58  & 0.10\\
 - & 900& 6.88 & -5.41 & 5.35 & 16.51& -0.31\\
 0.06&1000 & 6.59 & -5.24 & 5.40 & 15.69  & 0.03\\
 - & 900 & 9.23 & -5.43 & 5.42 &16.62 & -0.38\\
0.065& 1000 & 10. & -5.26 &  5.46 & 15.80 & -0.04\\
 -   & 900 & 13.9 & -5.46 &  5.47 & 16.72 & -0.44 \\
 \hline
\end{tabular}
\tablenotetext{*}{Synthetic photometry on the Johnson-Glass system as in AH95. On this system, the Hale spectrum of Gl229B gives J-H $= -0.115$, H-K $= -0.130$, and J-K $= -0.245$ for the brown dwarf.}
\end{table}

\onecolumn

\centerline{\bf Figure Captions}
\bigskip

\noindent{\bf Figure 1:} Synthetic flux distributions of brown dwarfs
with $\te$ between 1600 and 900~K, log~g=5.0 and assumed solar metallicity 
(solid) are compared to the latest broad band photometry (shown with error
bars) of Gl229B (Matthews \etal\, IAU Circular 6280).  A methane-less spectrum 
(dotted) is shown for one of the models (1000~K) to emphasize the position 
of these bands.  The photometric fluxes are normalized to each model at the 
peak of the J band flux.  Black body distributions of same effective 
temperatures are also shown (dashed) for comparison. 

\medskip
\noindent{\bf Figure 2:} The observed spectrum of Gl229B (thick) is 
compared to two $\te$ series of solar composition models of varying 
gravity.  The $\te=$ 1000~K sequence (upper panel) ranges from log~g= 
4.5 (dotted) to 5.8 (dashed), with the best fit reached for 5.3 (thin
full). The $\te=$ 900~K sequence (lower panel) ranges from log~g= 4.0 
(dotted) to 5.5 (dashed), with the best fit reached for 4.8 (thin full). 
Models and observations have been renormalized in each window to unity 
at the peak of the J bandpass.  Two methane systems not explicitely 
indicated on the plot are seen in this range at 1.6-1.8~$\mu$m and at 
2.2-2.5~$\mu$m.

\medskip
\noindent{\bf Figure 3:} Effective temperature as a function of time for 
the indicated masses in $\mso$.  The dash-dotted line corresponds to a 0.045 
$\mso$ model based on grey atmosphere including grain absorption and the 
dotted line gives the grainless grey counterpart.  The dashed horizontal 
lines give the $\te$ range determined by the present spectral analysis.

\medskip
\noindent{\bf Figure 4:} Gravity as a function of $\te$. The circles 
give the best fit deduced from the synthetic spectra analysis, with an
estimated theoretical error bar $\log \, g \approx \pm 0.2$. ($\teff$ = 1000K; 
$log \, g =$ 5.3 filled circle; $\teff$ = 900K; $log \, g =$ 4.8 open circle).


\end{sf}
\end{document}